\documentclass[a4paper,leqno,11pt]{amsart}
\pdfoutput=1

\usepackage{natbib,multirow,booktabs,color,subfigure,setspace,graphicx,amsaddr}
\usepackage[colorlinks=true,linkcolor=blue,citecolor=blue,urlcolor=blue]{hyperref}
\usepackage[left=1in,right=1in,top=1.4in,bottom=1.2in]{geometry}

\numberwithin{equation}{section}   

\newcommand{\mbs}[1]{\boldsymbol{#1}}
\newcommand{\mbf}[1]{\mathbf{#1}}

\newcommand{\veca}{\mathbf{a}}
\newcommand{\vecb}{\mathbf{b}}

\newcommand{\vecm}{\mathbf{m}}
\newcommand{\vecx}{\mathbf{x}}
\newcommand{\vecz}{\mathbf{z}}
\newcommand{\veczst}{\mathbf{z}^*}
\newcommand{\vecg}{\mathbf{g}}
\newcommand{\vech}{\mathbf{h}}
\newcommand{\vecq}{\mathbf{q}}
\newcommand{\vecW}{\mathbf{W}}
\newcommand{\vecy}{\mathbf{y}}

\newcommand{\vecsig}{\boldsymbol{\sigma}}
\newcommand{\matC}{\mathrm{C}}
\newcommand{\matR}{\mathrm{R}}
\newcommand{\matvar}{\Pi}
\newcommand{\sigatm}{\sigma_{\text{\tiny ATM}}}
\newcommand{\Tex}{T_{e}}

\newcommand{\grad}{\boldsymbol{\nabla}}

\begin{document}

\author[J. Choi]{Jaehyuk Choi}
\address{Peking University HSBC Business School}
\email{\href{mailto:jaehyuk@phbs.pku.edu.cn}{jaehyuk@phbs.pku.edu.cn}}

\author[S. Shin]{Sungchan Shin}
\address{Korea Advanced Institute of Science and Technology}
\email{\href{mailto:holypraise83@gmail.com}{holypraise83@gmail.com}}
\thanks{Address correspondence to Sungchan Shin,
	Department of Mathematical Sciences, KAIST, 335 Gwahangno, Yuseong-gu, Daejeon, Republic of Korea}

\title[Fast swaption pricing]{Fast swaption pricing in Gaussian term structure models}
\date{March 14, 2018}

\begin{abstract}
We propose a fast and accurate numerical method for pricing European
swaptions in multi-factor Gaussian term structure models. Our method
can be used to accelerate the calibration of such models to the
volatility surface. The pricing of an interest rate option in such a
model involves evaluating a multi-dimensional integral of the payoff
of the claim on a domain where the payoff is positive. In our method,
we approximate the exercise boundary of the state space by a
hyperplane tangent to the maximum probability point on the boundary
and simplify the multi-dimensional integration into an analytical
form. The maximum probability point can be determined using the
gradient descent method. We demonstrate that our method is superior to
previous methods by comparing the results to the price obtained by
numerical integration.
\end{abstract}

\keywords{Gaussian term structure model, volatility surface calibration, fast swaption pricing, swaption analytics}

\maketitle


\section{Introduction}
\label{sec:intro}

Swaptions, which are options on interest rate swaps, are the simplest
and most liquid option products traded in fixed income markets. From
practical and theoretical perspectives, swaptions are important
building blocks for more complicated claims, such as Bermudan callable
swaps. Swaptions are traded to hedge the volatility risk of such
exotic claims. Therefore, the parameters of a term structure model
must be calibrated to exactly reproduce the prices of the swaptions
observed in the market before they are used to price exotic
claims. However, the calibration process is typically a nonlinear
multi-dimensional root solving problem for which parameters must be
found using iterative methods. Therefore, it is critical to have a
fast and reliable method to price swaptions given a set of parameters
for a term structure model.

The most relevant studies on this topic are by \citet{SU} and
\citet{SP}.\footnote{Other original approaches have been proposed
  (e.g., \citet{Munk} and \citet{CDP}). These alternatives are
  dominated in terms of accuracy and computational cost. See
  \citet{SU} and \citet{SP} for details.}
Both studies provide a fast pricing method for the class of affine
term structure models (ATSM). \citet{SU} observe that the non-linear
exercise boundary for swaptions can be approximated by a
hyperplane. They compute the probability over the approximated domain
using the transform inversion method developed by \citet{DPS} and
\citet{MB}.  \citet{SP} derive an approximated stochastic differential
equation (SDE) for the underlying swap rate from full interest rate
dynamics, from which the swaption price is easily obtained. They
assume that the low variance martingale (LVM), which is typically the
ratio of the discount factors, is constant as time-zero
value. \citet{piterbarg_vol2_swapdyn} further refine the method by
improving the estimation of LVMs.  Because of its easy and intuitive
implementation, the \citet{SP} method has been favored by
practitioners. Considering that the method of freezing LVMs is
inspired by a similar method for pricing swaptions in the LIBOR Market
Model (LMM), their method is arguably the dominant swaption pricing
method for all classes of interest rates term structure
models. Although the \citet{SU} method appears to be equally
promising, it suffers from several drawbacks. First, because it lacks
explicit guidance in selecting the hyperplane, it fails to provide the
best hyperplane to minimize the error. Second, even for a given
hyperplane, the probability over the region must be computed under
different forward measures; there are as many measures as the number
of cash flows of the underlying swaptions.

This study demonstrates that the hyperplane approximation can be
significantly improved for the class of Gaussian term structure models
(GTSM). Using the analytical tractability of the GTSM, we can overcome
the two drawbacks mentioned above. In the GTSM, the probability
density function of the state is simply a multivariate Gaussian. In
other words, the GTSM is similar to the ATSM, where the transform
inversion is analytically solved. The knowledge of the density
function enables us to find the ÔbestÕ hyperplane to approximate the
non-linear boundary. We identify the point on the boundary with the
maximum probability density and determine the hyperplane tangent at
that point.

The accuracy of our approximation is better than the accuracy of
previous methods by several orders of magnitude, regardless of the
moneyness, expiry and tenor of the swaptions. Moreover, our method
does not sacrifice computational cost. The computational cost grows at
most linearly with the number of factors of the GTSM.  Although our
method is limited to the GTSM, which is a subset of the ATSM, it is
still a significant improvement for swaption calibration given the
indisputable importance of the GTSM among all term structure
models. Several previous term structure models are special cases of
the GTSM, e.g., \citet{HoLee}, \citet{Hull} and \citet{Vasicek}.
These models are still used by practitioners in their extended forms.
In the GTSM, we have the added benefit of being able to compare our
approximation to the exact swaption price. In contrast to the general
ATSM, where it is necessary to resort to Monte Carlo simulations, we
can obtain the exact price by combining the analytical result with
numerical integration. Thus, we can provide an accurate error
analysis.

The remainder of the paper is organized as follows. In Section 2, we
briefly review the Gaussian term structure model. Section 3 describes
the hyperplane approximation method and the exact swaption
pricing. Section 4 demonstrates the accuracy of our method and
compares it to previous methods.


\section{Multi-factor Gaussian term structure model}
\label{sec:model}

In this section, we review the important results of the GTSM. We will
define the scope of the GTSM and describe the preconditions for which
our approximation method is valid.  To simplify notation, define an
element-wise multiplication operator, $\circ$ , between vectors or
between a vector and a matrix by
\begin{gather}
  \mbf{a} \circ \mbf{b} = \mbf{b} \circ \mbf{a} = \big[\; a_j b_j\; \big]_{j}\;,\\
  \mathrm{M} \circ \mbf{a} = \mbf{a} \circ \mathrm{M} = \big[\; M_{jk} a_j\; \big]_{j,\,k}\;,\\
  \text{and}\quad \mathrm{M} \circ \mbf{a}^\top = \mbf{a}^\top \circ \mathrm{M} =
  \big[\; M_{jk} a_k\; \big]_{j,\,k}\;,
\end{gather}
where $\veca = [a_j]_j$ and $\vecb=[b_j]_j$ are $d\times 1$ vectors and
$\mathrm{M}=[M_{jk}]_{j,\,k}$ is a $d\times d$ matrix.

The GTSM in this study is a subclass of the Heath-Jarrow-Morton (HJM)
model class \citep{HJM}.  A general $d$-dimension HJM model starts
with the dynamics of the price of a zero-coupon bond. Let $P(t,T)$ be
the time-$t$ price of a zero-coupon bond maturing at $T$ and let the
SDE be defined as follows:
\begin{equation}
  \frac{dP(t,T)}{P(t,T)} = r(t)\,dt - \vecsig_P(t,T)^\top d\vecW^\beta(t),
\end{equation}
where $r(t)$ is the short rate process; $-\vecsig_P(t,T)$ is the
volatility vector; and $\vecW^\beta(t)$ is a $d$-dimensional Brownian
motion under the risk-neutral measure $Q^\beta$. The components of
$\vecW^\beta(t)$ are correlated with a correlation matrix $\matR(t) =
\left[\,\rho_{jk}(t)\,\right]_{j,\,k}$ where $\rho_{kk}(t)= 1$. If
$f(t,T)$ is the instantaneous forward rate (IFR) for time $T$ observed
at the current time $t$, we can write $P(t,T) = \exp\left(-\int_t^T
f(t,\,s) \;ds\right)$. An important result of the HJM model is
obtained by inserting this equation into the SDE for $P(t,T)$, to show
that \begin{equation} df(t,T) = \vecsig_f(t,T)^\top \vecsig_P(t,T)\,dt
  + \vecsig_f(t,T)^\top d\vecW^\beta(t),
\end{equation}
where the volatility of IFR $\vecsig_f(t,T)$ is as follows:
\begin{equation}
\vecsig_f(t,T) = \frac{\partial}{\partial T}\, \vecsig_P(t,T).
\end{equation}
Furthermore, the short rate process $r(t)$ is 
\begin{equation}
r(t) = f(t,t) = f(0,t) + \mbf{1}^\top \vecx(t),
\end{equation}
 for a $d \times 1$ state vector process $\vecx(t)$ with
$\vecx(0)=0$ and
\begin{equation}
 \vecx(t) = \int_0^t \vecsig_f(s,t) \circ \int_s^t \vecsig_f(s,u)\;du\,ds + \int_0^t \vecsig_f(s,t) \circ d\vecW^\beta(s).
\end{equation}
We can further simplify the result under the $t$-forward measure $Q^t$. Using the Girsanov theorem, we obtain
\begin{equation}
  d\vecW^\beta(s) = d\vecW^t(s) - \vecsig_P(s,t)ds = d\vecW^t(s) - \int_s^t \vecsig_f(s,u)\,du\;ds,
\end{equation}
where $\vecW^t(\cdot)$ is the Brownian motion under the $Q^t$ measure. The
processes for $f(s,t)$ and $\vecx(s)$ with respect to the time $s$
become driftless 
\begin{equation}
\label{eq:Tfwd}
df(s,t) = \vecsig_f(s,t)^\top d\vecW^t(s) \quad \text{and} \quad
\vecx(t) = \int_0^t \vecsig_f(s,t) \circ d\vecW^t(s).
\end{equation}
This result is consistent with the intuitive observation that $f(s,t)$
is a Martingale under the $Q^t$ measure with respect to time $s$.

An important result from \citet{HJM} is that, given the interest rate
curve $f(0,T)$ as an input to the model, the diffusion of the interest
rate curve is fully defined by specifying the volatility
$\vecsig_f(t,T)$. However, an HJM model typically imposes restrictions
on $\vecsig_f(t,T)$ because the process is generally path-dependent or
non-Markovian.

It is known that an HJM model is Markovian if and only if the
$\vecsig_f(t,T)$ is deterministic and separable in the form of
$\mathrm{G}(T)\vech(t)$ for a $d\times d$ matrix $\mathrm{G}(T)$ and a
$d\times 1$ vector $\vech(t)$~\citep{piterbarg_vol2}. Based on this
assumption, the IFR $f(t,T)$ and the state vector $\vecx(t)$ are
also Gaussian.

Although it is not a necessary condition for this study, a popular
choice for a separable form of $\vecsig_f$ is  one that causes the
short rate $r(t)$  to follow a mean-reverting Ornstein-Uhlenbeck process,
\begin{equation}
  d\vecx(t) = (\matvar(t)\mbf{1} - \mbs{\lambda}(t)\circ\vecx(t))dt + \vecsig_r(t)^\top d\vecW^\beta(t)
\end{equation}
for a deterministic mean reversion coefficient $\mbs{\lambda}(t)$,
a deterministic short rate volatility vector $\vecsig_r(t)$ and a drift matrix $\matvar(t)$.
This choice is equivalent to setting 
\begin{equation}
  \vecsig_f(t,T) = \vecsig_r(t) \circ \vecm(t,T),
\end{equation}
where $\vecm(t,T)$ is the exponential decay factor between time $t$ and $T$, defined by
\begin{equation}
  \vecm(t,T) = \left[\; e^{-\int_t^T \lambda_k(s)ds} \;\right]_k
  \quad \text{for} \quad \mbs{\lambda}(t) = \Big[\;\lambda_k(t)\;\Big]_k \;.
\end{equation}
The state $\vecx(t)$ is multivariate Gaussian. The drift matrix $\matvar(t)$ is
 the covariance matrix of  $\vecx(t)$. From Eq.~(\ref{eq:Tfwd}), we obtain
\begin{align}
\matvar(t) &= \int_0^t (\vecsig_f(s,t)\circ d\vecW^t(s))\,(\vecsig_f(s,t)\circ d\vecW^t(s))^\top \\
&= \int_0^t \vecsig_r(s) \circ \vecm(s,t) \circ
\matR(s) \circ \vecsig_r(s)^\top \circ \vecm(s,t)^\top ds \, \nonumber \\
&= \left[ \int_0^t \rho_{jk}(t) \sigma_{rj}(s) \sigma_{rk}(s) m_j(s,t) m_k(s,t) \; ds \right]_{j,\,k}. \nonumber
\end{align}

Finally, we can reconstruct the zero-coupon bond price at the future time $t$
using the Markovian state $\vecx(t)$ as 
\begin{equation}
\label{eq:recon1}
P(t,T) = \frac{P(0,T)}{P(0,t)}\exp\left( -\vecg(t,T)^\top \vecx(t)
- \frac12\, \vecg(t,T)^\top \matvar(t)\: \vecg(t,T)\right),
\end{equation}
where $\vecg(t,T) = \int_t^T \vecm(t,s)\,ds$. The volatility of
$P(t,T)$ is conveniently given as $\vecsig_P(t,T) =
-\vecsig_r(t)\circ\vecg(t,T)$; thus, $\vecg(t,T)$ is the risk loading
of the zero-coupon bond with respect to the short rate volatility.  It
should be noted that under the $t$-forward measure, $\vecx(t)$ has a
zero mean, and $E^{Q^t}\{ P(t,T) \} = P(0,T)/P(0,t)$. This result is
consistent with the fact that $P(s,T)/P(s,t)$ is a Martingale with
respect to time $s$ under the $Q^t$ measure.


\section{Swaption pricing method}
\label{sec:method}

Here, we derive the price of a swaption using the result from the
previous section.  Let us assume that the underlying swap of a
swaption begins at a forward time $T_0$, pays the fixed coupon $K$
(payer swap) on a payment schedule $\{T_1,\cdots T_m\}$, and receives
the floating interest rate, typically LIBOR.  The value of this
underlying swap, at time $t$, is
\begin{equation}
\text{V}(t) = P(t,T_0) - P(t,T_m) - \sum_{k=1}^m K P(t,T_k) \Delta_k,
\end{equation}
where $\Delta_k$ is the day count fraction for the $k$-th period,
$\Delta_k = T_k - T_{k-1}$. In general,
\begin{equation}
\text{V}(t) = \sum_{k=0}^m P(t,T_k)\, \text{CF}(T_k),
\end{equation}
for the  cash flow series $\text{CF}(T_k)$ at time $T_k$.

Let $T_e$ be the expiry of the swaption to enter into the underlying
swap paying the fixed coupon $K$ (the expiry $\Tex$ is typically two
business days before the start of the swap $T_0$).  Using the
reconstruction formula Eq.~(\ref{eq:recon1}), we can express the
future value of the swap as a function of the state $\vecx(\Tex)$.

We first decorrelate and normalize the state variable $\vecx(t)$ into
$\vecz(t)$ by $\vecx(t)=C(t)\;\vecz(t)$ for a Cholesky decomposition,
$\matC(t)$, of the covariance matrix $\matvar(t)$ that satisfies
$\matvar(t) = \matC(t)\,\matC(t)^\top$.  Then, the reconstruction
formula becomes
\begin{equation}
\label{eq:recon2}
P(t,T) = \frac{P(0,T)}{P(0,t)}\exp\left( -\veca(t,T)^\top\,\vecz(t)
- \frac12\, |\veca(t,T)|^2\right).
\end{equation}
where $\veca(t,T) = \matC(t)^\top\,\vecg(t,T)$.

We now have the value of the swap at the expiry $\Tex$ as a function
of the state $\vecz=\vecz(\Tex)$
\begin{align}
V(\vecz) = \frac1{P(0,\Tex)}\;\sum_{k=0}^m \text{DCF}_k\;
\exp\left(-\veca_k^\top \vecz - \frac12\, |\veca_k|^2 \right)
\end{align}
where $\text{DCF}_k$ is the discounted cash flow $\text{CF}(T_k)
P(0,T_k)$, and $\veca_k = \veca(\Tex,T_k)$. The price of
the payer swaption is the expectation of $V(\vecz)$ under the $T_e$-forward
measure,
\begin{equation}
\label{eq:payer}
C = P(0,\Tex)\;E^{Q^{\Tex}}\left\{\;\max(V(\vecz),0)\;\right\}.
\end{equation}
Similarly, the $\Tex$-forward price of the receiver swaption is  given by
\begin{equation}
P = P(0,\Tex)\;E^{Q^{\Tex}}\left\{\;\max(-V(\vecz),0)\;\right\}.
\end{equation}
The remainder of this study focuses on the payer swaption. The
receiver swaption can be determined from the put-call parity relation.


\subsection{Hyperplane approximation}

The evaluation of Eq.~(\ref{eq:payer}) involves a $d$-dimensional
integration. The difficulty lies in identifying the integration domain
$\Omega$, where the underlying swap has a positive value and the
boundary $\partial \Omega$ can be given as
\begin{equation}
  \Omega = \{\vecz:\;V(\vecz)\ge 0\}, \quad
  \partial \Omega = \{\vecz:\;V(\vecz)= 0\}.
\end{equation}
As described by \citet{SU}, we simplify the integration by
approximating the boundary $\partial \Omega$ as a hyperplane. However,
in this study, we substantially refine the original idea by providing
a systematic way to determine the ÔbestÕ hyperplane to use for this
approximation.

\begin{figure}[!h]
\centering
\includegraphics[width=0.75\textwidth]{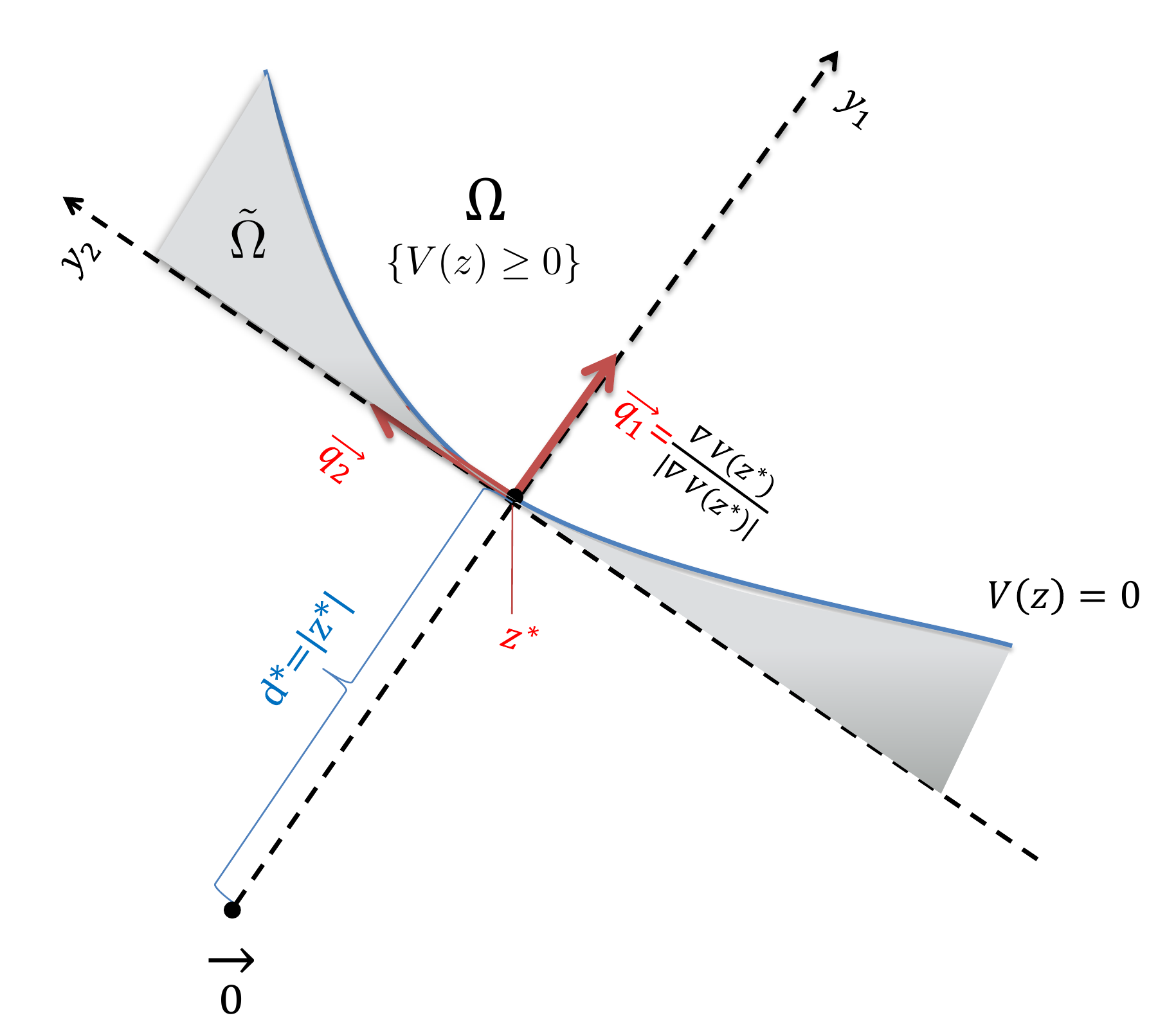}
\caption{Schematic of the hyperplane approximation method. The
  domain of state variables, $\Omega$, includes the area where the swaption payoff is in the money. We approximate
  the boundary of $\Omega$, $\partial \Omega$, as a hyperplane tangent
  ($y_1=0$) to the point $\veczst$ where the probability density is
  maximal on $\partial \Omega$ and integrate the payoff over the
  domain above the hyperplane ($y_1\ge 0$). Therefore, the integration
  over the shaded area is incorrectly added to our
  approximation. However, the error is extremely  small, as
  the swaption payoff is close to zero, and the probability density is
  small because of the choice of the tangent point, $\veczst$.
}
\label{fig:hyperplane}
\end{figure}

At the exercise boundary, we identify the state $\veczst$ with the
maximum probability density. Then, for the approximation, we use the
tangent plane to the boundary at $\veczst$, which is a systematic way
of choosing the hyperplane without ad-hoc rules based on
experience. This technique can be applied to any moneyness of
swaptions and any dimension of the GTSM. Because $\vecz$ has an
uncorrelated multivariate normal distribution, the probability density
decreases as a function of $|\vecz|$. Thus, $\veczst$ is also the
point with the shortest distance to the origin $\mbs{0}$ among the
points on the boundary $\partial\Omega$.  Furthermore, it follows that
$\veczst$ should be a scalar multiple of $\grad V(\veczst)$. We will
use this property to find $\veczst$. See Figure~\ref{fig:hyperplane}
for a geometric illustration.

The point $\veczst$ can be found numerically using the following iterative method.
One iteration consists of the following two steps: from $\vecz^{(i)}$ to
$\vecz^{(i+\frac12)}$ and from $\vecz^{(i+\frac12)}$ to $\vecz^{(i+1)}$.
\begin{description}
\item[(1)] First, apply a step of steepest descent method from $\vecz^{(i)}$:
\begin{equation}
  \vecz^{(i+\frac12)} = \vecz^{(i)} -
  \frac{ V(\vecz^{(i)}) }{|\grad V(\vecz^{(i)})|^2} \grad V(\vecz^{(i)})
\end{equation}
\item[(2)] Then, project $\vecz^{(i+\frac12)}$ onto the gradient direction
$\grad V(\vecz^{(i+\frac12)})$ by
\begin{equation}
\vecz^{(i+1)} = \frac{\grad V(\vecz^{(i+\frac12)})^\top \vecz^{(i+\frac12)}}{|\grad
  V(\vecz^{(i+\frac12)})|^2}\;\grad V(\vecz^{(i+\frac12)}).
\end{equation}
\end{description}
The convergence to the root is satisfied when the error
$V(\vecz^{(i+1)})$ is below a certain threshold; we use $10^{-13}$
in our study. 

It is difficult to prove with mathematical rigor that the iterative
scheme converges to a root for all possible parameterizations. However,
our method works without failure for any reasonable parameterization.
In the GTSM, the state variable is proportional to the interest rate as a
leading order, even allowing negative interest rates. Thus, it is
possible to find a state $\vecz$ for which the swap rate is equal to
any given (even negative) strike, which ensures that a boundary
$\partial \Omega$ always exists and  is close to a hyperplane.
This iterative root-finding process uses the most time in the entire
computation of the swaption price because the rest of the
computation is analytical. A reasonably calibrated GTSM 
converges to the root $\veczst$ quickly, typically within 7 iterations
starting from the origin. It should be noted that the number of
iterations does not increase with the dimension $d$, although the
computational cost may increase due to the increasing number of
components. Overall, the computation cost  increases linearly, not exponentially, with
the dimension $d$.

Now, we can simplify the evaluation of Eq.~(\ref{eq:payer}). Let $\vecq_1 =
\grad V(\veczst)/|\grad V(\veczst)|$ be the unit vector in the
direction of $\grad V(\veczst)$. We express the state vector as follows:
\begin{equation}
  \vecz = y_1\,\vecq_1 + \cdots + y_d\,\vecq_d
\end{equation}
in a Cartesian coordinate $\vecy=(y_1,\cdots,y_d)$ with the standard
basis $\{\vecq_1,\cdots,\vecq_d\}$.  The unspecified unit vectors
$\vecq_2,\cdots,\vecq_d$ can be chosen arbitrarily as long as the
matrix of unit vectors $\mathrm{Q} = \big[\, \vecq_1, \cdots,
  \vecq_d\,\big]$ forms an orthogonal matrix.

Using the above property, we select a scalar $d^*$ and vectors
$\vecb_k$ such that
\begin{equation}
\veczst = d^* \, \vecq_1, \qquad \veca_k = \mathrm{Q} \,\vecb_k
\end{equation}
The approximated domain $\tilde{\Omega}$ is 
\begin{equation}
\tilde{\Omega} = \{\vecz:\; \vecq_1^\top (\vecz - \veczst)\ge 0\}
= \{\vecy:\; y_1 \ge d^*\}
\end{equation}
and the value of the swap becomes
\begin{equation}
  V(\vecz) = \frac1{P(0,\Tex)}\;\sum_{k=0}^m \text{DCF}_k 
  \exp\left(-\frac12\, |\vecb_k|^2 - \vecb_k^\top \,\vecy\right)
\end{equation}
When integrating on the $\vecy$ coordinate, $y_1$ is the only axis on which
the integration is non-trivial. The rest of the dimensions integrate to unity.
We obtain the swaption price in analytic form as follows:
\begin{align}
\label{eq:integ1}
C &= P(0,\Tex)\,\int_{\Omega} V(\vecz) f(\vecz)\; d\vecz \approx
P(0,\Tex)\,\int_{\tilde{\Omega}} V(\vecy) f(\vecy)\; d\vecy\\
&= \sum_{k=0}^m \text{DCF}_k \int_{-\infty}^\infty\!\! dy_2\cdots dy_d \int_{d^*}^\infty\!\! dy_1
\;\;e^{-\frac12\, |\vecb_k|^2 - \sum_{j=1}^d b_{kj}y_j } n(y_1)\cdots n(y_d) \nonumber\\
&= \sum_{k=0}^m \text{DCF}_k \int_{d^*}^\infty\!\! dy_1 \;\; e^{-\frac12\, b_{k1}^2 - b_{k1} y_1} n(y_1)\nonumber\\
&= \sum_{k=0}^m \text{DCF}_k \;N(-b_{k1}-d^*) = \sum_{k=0}^m \text{DCF}_k \;N(-(\veca_k+\veczst)^\top \vecq_1), \nonumber
\end{align}
where $f(\cdot)$ is the probability density function of the
multivariate normal distribution, and $n(\cdot)$ and $N(\cdot)$ are
the probability and cumulative density functions of the univariate
normal distribution, respectively.

\subsection{Exact pricing method}
Although it is computationally demanding, we can combine numerical
integration and analysis to price the swaption precisely and measure
the accuracy of the hyperplane approximation. The integration is
performed on the $\vecy$ coordinate, i.e., in the hyperplane
approximation. However, in the exact method, we numerically determine
the distance to the boundary $d$ for each given $(d\!-\!1)$-tuple
$(y_2,\cdots,y_d)$:
\begin{equation}
  \Omega=\{\vecy:\; y_1 \ge d(y_2,\cdots,y_d)\}.
\end{equation}
The root finding for $d$ can be determined using the Newton-Raphson
method in one-dimension.  The integration is performed analytically
for $y_1$ and numerically for the rest of the dimensions:
\begin{align}
\label{eq:integ2}
C &= P(0,\Tex)\,\int_{\Omega} V(\vecz) f(\vecz)\; d\vecz =
P(0,\Tex)\,\int_{\Omega} V(\vecy) f(\vecy)\; d\vecy\\
&= \sum_{k=0}^m \text{DCF}_k \int_{-\infty}^\infty \!\! dy_2 \cdots dy_d \int_d^\infty\!\! dy_1 \;
e^{-\frac12\, |\vecb_k|^2 - \sum_{j=1}^db_{kj}y_j  } n(y_1)\cdots n(y_d)  \nonumber\\
&= \sum_{k=0}^m \text{DCF}_k\int_{-\infty}^\infty\!\! dy_2\cdots dy_d \;\;
N(-b_{k1}-d(y_2,\cdots, y_d)) n(y_2+b_{k2})\cdots n(y_d+b_{kd})  \nonumber
\end{align}
We can use the finite difference method for the numerical integration.

It should be noted that the error from the hyperplane approximation is due to the difference
between the integrands of Eq.~(\ref{eq:integ1}) and Eq.~(\ref{eq:integ2}):
\begin{equation}
\label{eq:error}
E(y_2, \cdots, y_d) = \sum_{k=0}^m \left( N(-b_{k1}-d^*) - N(-b_{k1}-d(\cdots)) \right)
n(y_2+b_{k2})\,\cdots\,n(y_d+b_{kd})
\end{equation}
We will examine this error through examples in the next section. It is interpreted as the error density because the error in the swaption price is 
\begin{equation}
  \text{Price Error}= \int_{-\infty}^\infty \!\! dy_2 \cdots dy_d\; E(y_2, \cdots, y_d)
\end{equation}


\section{Approximation quality and comparison to other methods}
\label{sec:result}

To examine the quality of the proposed hyperplane approximation method
for swaption pricing, we apply it to three sets of examples, shown in
Table~\ref{tab1} to Table~\ref{tab3}. The first two examples use
different parameter sets in a two-factor GTSM calibrated to realistic
swaption volatility surfaces in the least-square sense.  We select two
contrasting market conditions in the shapes of the yield curve and the
volatility surface to test our approximation in diverse market
environments.  In the first example, the market sees high uncertainty
in the short-term interest rate, and the yield curve is flat at $5\%$
at time 0. In the second example, the market sees high uncertainty in
the long-term interest rate, and the interest rate curve increases
steeply from the $0\%$ short-term interest rate, most likely because
of monetary policies.

To calibrate the surface as closely as possible, we use a
piece-wise-constant term structure for volatility and a mean reversion
structure for the first factor with Parameter Sets 1 and 2. The
parameters for the second factor are specified through the constant
volatility ratio $\sigma_2/\sigma_1$ and the constant mean reversion
difference $\lambda_2 - \lambda_1$. This structure allows the
instantaneous correlation between $f(t,T_1)$ and $f(t,T_2)$ to be
stationary~\citep{piterbarg_vol2_corr}.

For the third example, we reuse the three-factor GTSM parameter set
from \citet{CDP}.  This parameter set was also used by \citet{SP} to
compare their result to those of \citet{CDP}.


\begin{table}[!h]
\caption{Parameter Set 1: A two-factor Gaussian model calibrated to a
  volatility surface, where the swaptions on the shorter tenor swaps
  are relatively expensive. The current forward rate curve is assumed to
  be flat at $5\%$.}
\label{tab1}
\begin{center}
\begin{tabular}{|c|c|c|c|c|c|c|}
\hline
 Time(year)    &   0 $\sim$ 0.25    &   $\sim$  0.5    &   $\sim$ 1     &   $\sim$  2  &   $\sim$  5 &   $\sim$\\
\hline
Volatility($\sigma_1$)  & 0.030 & 0.024 & 0.024 & 0.022 & 0.018 & 0.012\\
\hline
\hline

  Time(year)  &  \multicolumn{2}{c|}{ 0 $\sim$ 5}   & \multicolumn{2}{c|}{ $\sim$ 10 }    &  \multicolumn{2}{c|}{$\sim$}\\
\hline

Mean reversion($\lambda_1$)  & \multicolumn{2}{c|}{  0.115 }   & \multicolumn{2}{c|}{  0.073 }   & \multicolumn{2}{c|}{ 0.029 }\\
\hline
\hline
$\sigma_2/\sigma_1$ & \multicolumn{6}{c|}{ 1.05 }\\
\hline
$\lambda_2 -\lambda_1$ & \multicolumn{6}{c|}{ 0.27 }\\
\hline
$\rho_{12}$ & \multicolumn{6}{c|}{ -77\% }\\
\hline
$f(0,t)$ & \multicolumn{6}{c|}{ 5\% }\\
\hline
\end{tabular}
\end{center}
\end{table}


\begin{table}[!h]
\caption{Parameter Set 2: A two-factor Gaussian model calibrated to a
  volatility surface where the swaptions on the longer tenor swaps are
  relatively expensive. The current forward rate curve is assumed to
  increase steeply from $0\%$ to $6\%$.}
\label{tab2}
\begin{center}

\begin{tabular}{|c|c|c|c|c|c|c|}
\hline

 Time(year)    &   0 $\sim$ 0.25    &   $\sim$  0.5    &  $\sim$ 1     &   $\sim$  2  &   $\sim$  5 &   $\sim$\\
\hline

Volatility($\sigma_1$)  & 0.020 & 0.014 & 0.013 & 0.012 & 0.01 & 0.009\\
\hline
\hline

  Time(year)  &  \multicolumn{2}{c| }{ 0 $\sim$ 5}   & \multicolumn{2}{c|}{ $\sim$ 10 }    &  \multicolumn{2}{c|}{ $\sim$ }\\
\hline

Mean reversion($\lambda_1$)  & \multicolumn{2}{c|}{ -0.051 }   & \multicolumn{2}{c|}{ 0.059 }   & \multicolumn{2}{c|}{ 0.017 }\\
\hline
\hline
$\sigma_2/\sigma_1$ & \multicolumn{6}{c|}{ 1.05 }\\
\hline
$\lambda_2 -\lambda_1$ & \multicolumn{6}{c|}{ 0.27 }\\
\hline
$\rho_{12}$ & \multicolumn{6}{c|}{ -77\% }\\
\hline
$f(0,t)$ & \multicolumn{6}{c|}{ 6\%$\times(1-e^{-t/10})$ }\\
\hline

\end{tabular}
\end{center}
\end{table}


\begin{table}[!h]
\caption{Parameter Set 3: A three-factor Gaussian model from \citet{CDP} and \citet{SP}}
\label{tab3}
\begin{center}
\begin{tabular}{|ccc|ccc|ccc|c|}
\hline
$\sigma_1$  & $\sigma_2$  & $\sigma_3$  & $\lambda_1$ & $\lambda_2$ & $\lambda_3$  &   $\rho_{12}$ & $\rho_{13}$ & $\rho_{23}$ & $f(0,t)$\\
\hline
0.010  & 0.005  & 0.002  & 1.0  &  0.2  &  0.5  &   -20\%  & -10\%  & 30\% & 5.5\%\\
\hline
\end{tabular}
\end{center}
\end{table}


\begin{table}[p]
\caption{Prices and errors of the hyperplane approximation with
  Parameter Set 1 in basis point units. Relative pricing errors,
  calculated as fractions of exact prices, are in parentheses.}
\label{tab4}
\begin{small}
\begin{center}
\begin{tabular}{cccccc}
\toprule
option & \multicolumn{5}{c}{swap maturity}\\
expiry   &  1  &  2  &  5  &  10 & 30\\ \midrule

ATM  & \multicolumn{5}{l}{$K = F$}\\

1 & 54.54 (-9.1E-12) & 100.91 (-2.4E-09) & 213.22 (-1.0E-06) &  346.39 (-2.5E-05) & 572.21 (-5.0E-04)\\

2 & 65.39 (-9.8E-12) & 122.63 (-2.6E-09) & 264.82 (-1.1E-06) &  435.98 (-2.8E-05) & 729.43 (-5.9E-04)\\

5 & 71.55 (-5.0E-12) & 137.82 (-1.1E-09) & 308.60 (-5.3E-07) &  525.22 (-1.8E-05) & 898.62 (-4.2E-04)\\

10 & 62.45 (-9.5E-13) & 122.06 (-2.3E-10) & 283.69 (-1.4E-07) & 493.42 (-5.6E-06) & 842.60 (-1.5E-04) \\

20 & 	50.97 (-6.0E-13) & 99.08 (-1.4E-10) & 227.00 (-9.0E-08) &  389.08 (-3.8E-06) & 651.58 (-1.1E-04)\\ \midrule

ITM &  \multicolumn{5}{l}{$K = F - \sigatm\sqrt{\Tex}$}\\

1 & 147.97 (-2.2E-12) & 273.77 (-7.6E-10) & 578.43 (-3.4E-07) & 939.35 (-8.9E-06) & 1547.59 ( -1.9E-04)\\

2 & 177.39 (-2.6E-12) & 332.63 (-6.4E-10) & 718.27 (-2.9E-07) & 1181.97 (-7.9E-06) & 1970.51 (-1.9E-04)\\

5 & 194.04 (-4.9E-13) & 373.76 (-1.9E-10) & 836.80 (-9.6E-08) & 1423.69 (-3.4E-06)  & 2423.75 (-9.5E-05)\\

10 & 169.34 (2.4E-13) & 330.98 (-3.0E-11) & 769.25 (-1.9E-08) &    1337.53 (-7.5E-07) & 2269.43 ( -2.6E-05) \\ 

20 & 138.14 (6.1E-13) & 268.53 (-6.7E-12) & 615.15 (-4.9E-09) & 1053.67 (-2.3E-07)  & 1749.22 (-1.2E-05)\\ \midrule

OTM & \multicolumn{5}{l}{$K = F + \sigatm\sqrt{\Tex}$} \\

1 & 11.51 (-7.7E-12) & 21.31 (-2.4E-09) & 45.09 (-9.6E-07) & 73.60 (-2.3E-05) & 125.63 (-4.3E-04) \\

2 & 13.84 (-1.1E-11) & 25.97 (-2.9E-09) & 56.15 (-1.2E-06) & 93.00 (-2.9E-05)  & 162.31 (-5.7E-04)\\

5 & 15.20 (-5.8E-12) & 29.28 (-1.5E-09) & 65.66 (-6.7E-07) & 112.24 (-2.2E-05)  & 203.21 (-4.8E-04)\\

10 & 13.29 (-1.5E-12) & 25.97 (-3.3E-10) & 60.34 (-2.0E-07) & 105.39 (-8.0E-06)  &193.21 (-1.9E-04) \\ 

20 & 10.92 (-6.2E-13) & 21.22 (-2.4E-10)  & 48.67 (-1.6E-07)  & 84.12 (-6.5E-06)  & 154.65 (-1.6E-04) \\ \bottomrule
\end{tabular}
\end{center}
\end{small}
\end{table}


\begin{table}[p]
\label{tab5}
\caption{Implied normal volatilities and errors of the hyperplane
  approximation with Parameter Set 1 in daily basis point units.
  Relative volatility errors, calculated as fractions of exact
  volatilities, are in parentheses.}
\begin{small}
\begin{center}
\begin{tabular}{cccccc}
\toprule
option &\multicolumn{5}{c}{swap maturity}\\
expiry   & 1 & 2 & 5 & 10 & 30\\ \midrule

ATM & \multicolumn{5}{l}{$K = F$} \\		

1 & 9.45 (-1.6E-12) & 8.96 (-2.1E-10) & 8.14 (-3.8E-08) & 7.44 (-5.4E-07) & 6.22 (-5.4E-06)\\

2 & 8.40 (-1.3E-12) & 8.07 (-1.7E-10) & 7.50 (-3.0E-08) & 6.94 (-4.4E-07) & 5.88 (-4.7E-06)\\

5 & 6.74 (-4.7E-13) & 6.66 (-5.5E-11) & 6.41 (-1.1E-08) & 6.13 (-2.1E-07) & 5.32 (-2.5E-06) \\

10 & 5.34 (-8.3E-14) & 5.35 (-1.0E-11) & 5.35 (-2.7E-09) & 5.23 (-5.9E-08) & 4.52 (-8.1E-07) \\ 

20 &	5.08 (-6.0E-14) & 5.06 (-6.9E-12) & 4.9 (-2.0E-09) & 4.81 (-4.7E-08) & 4.08 (-6.9E-07) \\ \midrule

ITM & \multicolumn{5}{l}{$K = F - \sigatm\sqrt{\Tex}$} \\

1 & 9.41 (-6.2E-13) &   8.92 (-1.1E-10) & 8.11 (-2.1E-08) & 7.39 (-3.2E-07) & 6.11 (-3.5E-06) \\

2 & 8.36 (-5.6E-13) &   8.03 (-7.0E-11) &   7.46 (-1.4E-08) &   6.89 (-2.1E-07) & 5.74 (-2.6E-06) \\

5 & 6.70 (-8.0E-14) & 6.62 (-1.5E-11) & 6.37 (-3.3E-09) &   6.09 (-6.5E-08)  & 5.15 (-9.6E-07)\\

10 & 5.31 (3.6E-14) & 5.31 (-2.2E-12) & 5.31 (-5.8E-10) & 5.19 (-1.3E-08)  & 4.36 (-2.4E-07)\\ 

20 &	5.04 (9.5E-14) & 5.02 (-5.6E-13) & 4.94 (-1.8E-10) & 4.75 (-4.8E-09) & 3.86 (-1.3E-07) \\ \midrule
OTM & \multicolumn{5}{l}{$K = F + \sigatm\sqrt{\Tex}$} \\

1 & 9.48 (-2.2E-12) &   8.99 (-3.5E-10) & 8.18 (-6.0E-08) & 7.48 (-8.2E-07) &6.33	(-7.6E-06) \\

2 & 8.44 (-2.3E-12) &   8.11 (-3.1E-10) & 7.54 (-5.4E-08) & 6.99 (-7.6E-07)  & 6.01 (-7.4E-06)\\

5 & 6.78 (-9.0E-13) &   6.69 (-1.2E-10) &   6.45 (-2.3E-08) &   6.18 (-4.2E-07) & 5.47 (-4.6E-06) \\

10 & 5.37 (-2.1E-13) &  5.38 (-2.3E-11) &   5.38 (-6.3E-09) & 5.27 (-1.4E-07) & 4.67 (-1.6E-06) \\

20 &	5.12 (-1.0E-13) & 5.11 (-2.0E-11) & 5.03 (-5.7E-09) &	4.86 (-1.3E-07) & 4.26 (-1.5E-06)\\ \bottomrule
\end{tabular}
\end{center}
\end{small}
\end{table}


\begin{table}
\label{tab6}
\begin{small}
\caption{Prices and errors of the hyperplane approximation with
  Parameter Set 2 in basis point units. Relative pricing errors,
  calculated as fractions of exact prices, are in parentheses.}
\begin{center}
\begin{tabular}{cccccc}
\toprule
option &\multicolumn{5}{c}{swap maturity}\\
expiry   & 1 & 2 & 5 & 10 &30\\ \midrule

ATM & \multicolumn{5}{l}{$K = F$}\\

1 & 41.00 (-1.3E-13) & 84.22 (-4.9E-11) & 236.50 (-8.3E-08) & 459.17 (-4.8E-06)  & 835.96 (	-3.4E-04)\\

2 & 54.50 ( 1.7E-14) & 113.43 (-1.1E-10) & 307.97 (-1.2E-07) & 570.78 (-6.3E-06) &1019.81(-4.1E-04) \\

5 & 82.63 (-6.6E-13) & 162.31 (-1.1E-10) & 377.36 (-7.5E-08) & 663.98 (-3.9E-06) &1153.94 (-2.0E-04)\\

10 & 76.25 (-7.3E-13) & 150.19 (-7.7E-11) & 355.25 (-5.9E-08) & 629.49 (-3.0E-06) &1087.28 (-1.3E-04)\\

20 &	62.94 (-2.0E-13) & 122.64	(-8.3E-11) & 282.57 (-6.2E-08) & 487.73 (-3.0E-06) &	823.07 (-1.1E-04)\\ \midrule

ITM & \multicolumn{5}{l}{$K = F - 0.5\;\sigatm\sqrt{\Tex}$}\\

1 & 71.67 ( 4.3E-13) & 147.21 (-1.3E-11) & 413.65 (-3.0E-08) & 803.29 (-2.3E-06) &1459.50 (-2.1E-04)\\

2 & 95.25 (8.0E-13) & 198.25 (-2.6E-11) & 538.50 (-3.9E-08) & 997.95 (-2.8E-06) &1778.11 (-2.4E-04)\\

5 & 144.32 (1.2E-12) & 283.51 (-3.1E-11) & 659.05 (-2.8E-08) & 1159.42 (-1.7E-06) & 2007.45 (-1.1E-04)\\

10 & 133.17 (1.2E-12) & 262.31 (-3.2E-11) & 620.49 (-2.6E-08) & 1099.40 (-1.4E-06) &1890.57 (-7.1E-05) \\ 

20 &	109.86 (-8.7E-14) & 214.08 (-3.6E-11) &  493.27	 (-2.7E-08) &  851.14	(-1.3E-06) & 1427.93	(-5.6E-05) \\ \midrule

OTM & \multicolumn{5}{l}{$K = F + 0.5\;\sigatm\sqrt{\Tex}$}\\

1 & 20.38 (-4.9E-13) & 41.84 (-9.8E-11) & 117.28 (-1.4E-07) & 227.51 (-6.9E-06) & 417.10 (-4.0E-04) \\

2 & 27.10 (4.3E-14) & 56.40 (-2.1E-10) & 152.87 (-2.0E-07) & 283.39 (-9.5E-06) & 511.15 (-5.1E-04)\\

5 & 41.17 (-6.9E-13) & 80.87 (-1.9E-10) & 188.08 (-1.2E-07) & 331.17 (-5.8E-06) & 582.87 (-2.6E-04)\\

10 & 38.01 (-8.3E-13) & 74.86 (-1.2E-10) &  177.02 (-8.7E-08) & 313.74 (-4.3E-06) & 549.87 (-1.7E-04)\\ 

20 &	31.42 (0.0E+00) & 61.23 (-1.3E-10) &	141.08 (-9.2E-08) & 243.76 (-4.4E-06) &  419.20	 (-1.5E-04) \\ \bottomrule
\end{tabular}
\end{center}
\end{small}
\end{table}


\begin{table}
\caption{Implied normal volatilities and errors of the hyperplane
  approximation with Parameter Set 2 in daily basis point units.
  Relative volatility errors, calculated as fractions of exact
  volatilities, are in parentheses.}
\label{tab7}
\begin{small}
\begin{center}
\begin{tabular}{cccccc }
\toprule
option &\multicolumn{5}{c}{swap maturity}\\

expiry & 1 & 2 & 5 & 10 & 30 \\ \midrule

ATM & \multicolumn{5}{l}{$K = F$}\\

1 & 6.57 (-2.1E-14) & 6.78 (-4.0E-12) & 7.82 (-2.8E-09) & 8.11 (-8.5E-08) &  7.07	(-2.9E-06) \\

2 & 6.23 (1.8E-15) & 6.54 (-6.4E-12) & 7.33 (-2.8E-09) & 7.31 (-8.1E-08) & 6.32	(-2.6E-06) \\

5 & 6.34 (-5.0E-14) & 6.32 (-4.2E-12) & 6.16 (-1.2E-09) & 5.93 (-3.5E-08)  &5.14	(-9.1E-07) \\

10 & 4.89 (-4.6E-14) & 4.91 (-2.5E-12) & 4.95 (-8.2E-10) & 4.89 (-2.3E-08) & 4.35 (-5.3E-07 )\\ 

20 &	4.57 (-1.4E-14)	& 4.57 (-3.1E-12) & 4.55 (-9.9E-10) &	4.47 (-2.7E-08) 	& 4.00 (-5.4E-07)\\ \midrule

ITM & \multicolumn{5}{l}{$K = F - 0.5\;\sigatm\sqrt{\Tex}$}\\

1 & 6.56 (7.6E-14) & 6.77 (-1.2E-12) & 7.82 (-1.1E-09) & 8.11 (-4.6E-08) & 7.04	(-2.0E-06) \\

2 & 6.22 (1.0E-13) & 6.53 (-1.7E-12) & 7.33 (-1.1E-09) & 7.30 (-4.0E-08) & 6.29	(-1.7E-06)\\

5 & 6.33 (1.1E-13) & 6.30 (-1.4E-12) & 6.14 (-5.2E-10) & 5.91 (-1.8E-08) & 5.08	(-5.7E-07)\\

10 & 4.88 (9.0E-14) & 4.90 (-1.2E-12) & 4.94 (-4.1E-10) & 4.88 (-1.2E-08) & 4.30 (-3.2E-07)\\  

20 &	4.55 (-7.1E-15) & 4.55 (-1.5E-12) & 4.53 (-4.9E-10) &	4.44	(-1.4E-08) & 3.94 (-3.1E-07) \\ \midrule

OTM & \multicolumn{5}{l}{$K = F + 0.5\;\sigatm\sqrt{\Tex}$} \\

1 & 6.57 (-8.8E-14) & 6.79 (-8.9E-12) & 7.82 (-5.1E-09) &   8.10 (-1.4E-07) & 7.09 (-3.9E-06) \\

2 & 6.24 (6.2E-15) & 6.55 (-1.4E-11) & 7.34 (-5.5E-09) & 7.31 (-1.4E-07) & 6.36	(-3.6E-06) \\

5 & 6.36 (-6.2E-14) & 6.34 (-8.4E-12) & 6.17 (-2.2E-09) & 5.95 (-5.9E-08) & 5.19	(-1.3E-06) \\

10 & 4.90 (-6.3E-14) & 4.93 (-4.3E-12) & 4.96 (-1.4E-09) & 4.91 (-3.8E-08) & 4.40 (-7.7E-07)\\

20 &	4.59 (0.0E+00) & 4.59 (-5.3E-12) & 4.57 (-1.7E-09) &	4.49	(-4.5E-08) & 4.06 (-8.1E-07)\\ \bottomrule
\end{tabular}
\end{center}
\end{small}
\end{table}


\begin{table}
\caption{Prices and errors of the hyperplane approximation with
  Parameter Set 3 in basis point units. Relative pricing errors,
  calculated as fractions of exact prices, are in parentheses.}
\label{tab8}
\begin{small}
\begin{center}
\begin{tabular}{cccccc}
\toprule
option &\multicolumn{5}{c}{swap maturity}\\
expiry & 1 & 2 & 5 & 10 & 30\\ \midrule

ATM & \multicolumn{5}{l}{$K = F$}\\

1 & 20.65 (-5.9E-09) & 32.91 (-1.5E-07) & 53.27 (-1.5E-06) & 65.95 (-2.3E-06) & 70.86	(-2.6E-06)\\

2 & 23.46 (-9.0E-09) & 38.38 (-1.9E-07) & 63.98 (-1.5E-06) & 79.92 (-2.0E-06) & 86.07	(-2.1E-06) \\

5 & 23.45 (-9.2E-09) & 39.24 (-1.6E-07) & 66.99 (-1.1E-06) & 84.25 (-1.4E-06) & 90.90 (-1.4E-06)\\

10 & 18.69 (-7.2E-09) & 31.45 (-1.2E-07) & 53.97 (-8.1E-07) & 68.00 (-1.0E-06) & 73.40 (-1.0E-06)\\ 

20 &	10.85 (-4.1E-09) & 18.28 (-7.1E-08) & 31.40 (-4.6E-07) & 39.57 (-5.8E-07) & 42.72 (-5.8E-07)\\ \midrule

ITM & \multicolumn{5}{l}{$K = F - 2\;\sigatm\sqrt{\Tex}$}\\
1 & 103.93 (-6.33E-10) & 165.66 (-1.64E-08) & 268.15	 (-1.71E-07) & 331.94 (-2.70E-07) & 356.59 (-3.08E-07) \\
2 & 118.12 (-9.08E-10) & 193.20	 (-1.97E-08) & 322.08 (-1.64E-07) & 402.25 (-2.29E-07) & 433.10 (-2.46E-07) \\
5 & 118.05 (-8.72E-10) &  197.56 (-1.64E-08) & 337.19 (-1.17E-07) & 424.04 (-1.50E-07) & 457.36 (-1.56E-07) \\
10 & 	94.07 (-6.65E-10) & 158.29 (-1.21E-08) & 271.68 (-8.40E-08) & 342.24 (-1.07E-07) & 369.31 (-1.11E-07) \\
20 & 	54.64 (-3.83E-10) & 92.02 (-6.95E-09) & 158.06 (-4.81E-08) & 199.16 (-6.14E-08) & 214.93 (-6.37E-08) \\ \midrule

OTM & \multicolumn{5}{l}{$K = F + 2\;\sigatm\sqrt{\Tex}$}\\
1 &	0.45(	-9.78E-10) & 0.72 (-2.35E-08) & 1.17 (-2.27E-07) &1.48 (-3.52E-07) & 1.65 (-3.87E-07) \\
2 & 	0.51 (-1.53E-09) & 0.84 (-3.07E-08) & 1.41 (-2.37E-07) & 1.81 (-3.23E-07)	& 2.06 (-3.33E-07) \\
5 &	0.51(-1.63E-09) & 0.86 (-2.82E-08) & 1.49 (-1.84E-07) & 1.94 (-2.32E-07) & 2.23 (-2.28E-07) \\
10 &	0.41 (-1.28E-09) & 0.69 (-2.15E-08) & 1.20 (-1.36E-07) & 1.57 (-1.70E-07) & 1.82 (-1.67E-07) \\
20 &	0.24 (-7.44E-10) & 0.40 (-1.24E-08) & 0.70 (-7.85E-08) & 0.91 (-9.79E-08) & 1.06 (-9.56E-08) \\ \bottomrule
\end{tabular}
\end{center}
\end{small}
\end{table}


\begin{table}
\caption{Implied normal volatilities and errors of the hyperplane
  approximation with Parameter Set 3 in daily basis point units.
  Relative volatility errors, calculated as fractions of exact volatilities, are in parentheses.}
\label{tab9}
\begin{small}
\begin{center}
\begin{tabular}{cccccc } \toprule

option &\multicolumn{5}{c}{swap maturity}\\
expiry &  1  &  2  & 5 & 10 & 30 \\ \midrule	

ATM & \multicolumn{5}{l}{$K = F$}\\

1 & 3.61 (-1.0E-09) & 2.95 (-1.3E-08) & 2.07 (-5.7E-08) & 1.46 (-5.1E-08) & 0.82	(-3.0E-08) \\

2 & 3.06 (-1.2E-09) & 2.57 (-1.2E-08) & 1.85 (-4.3E-08) & 1.32 (-3.3E-08) & 0.74	(-1.8E-08)\\

5 & 2.27 (-8.9E-10) & 1.96 (-8.2E-09) & 1.45 (-2.4E-08) & 1.03 (-1.7E-08) & 0.58	(-9.0E-09)\\

10 & 1.69 (-6.5E-10) & 1.46 (-5.7E-09) &    1.08 (-1.6E-08) & 0.78 (-1.2E-08) & 0.44 (-6.1E-09)\\

20 & 	1.20	(-4.6E-10) & 1.04(-4.0E-09) & 0.77 (-1.1E-08) & 	0.55	(-8.1E-09) & 0.31 (-4.3E-09)\\  \midrule

ITM & \multicolumn{5}{l}{$K = F - 2\;\sigatm\sqrt{\Tex}$}\\
1 &	3.60  (-8.3E-10) & 2.94 (-1.1E-08) &	2.06 (-5.0E-08) & 1.45 (-4.6E-08) & 0.81 (-2.8E-08) \\
2 &	3.04 (-8.9E-10) & 2.56 (-9.9E-09) & 1.84 (-3.6E-08) & 1.30 (-2.9E-08) & 0.73 (-1.7E-08) \\
5 & 	2.26 (-6.4E-10) & 1.95 (-6.1E-09) & 1.44 (-1.9E-08) & 1.02 (-1.4E-08) & 0.57 (-8.2E-09) \\
10 &	1.68 (-4.5E-10) & 1.45 (-4.2E-09) & 1.08 (-1.3E-08) & 	0.77 (-9.5E-09) & 0.43 (-5.5E-09) \\
20 &	1.19 (-3.2E-10) & 1.03 (-3.0E-09) & 0.77 (-9.0E-09) &	0.55 (-6.7E-09) & 0.30 (-3.8E-09) \\ \midrule

OTM & \multicolumn{5}{l}{$K = F + 2\;\sigatm\sqrt{\Tex}$}\\
1 &	3.62 (-1.2E-09) & 2.96 (-1.5E-08) & 2.08 (-6.4E-08) & 1.47 (-5.6E-08) & 0.83 (-3.1E-08) \\
2 &	3.07 (-1.5E-09) & 2.58 (-1.5E-08) & 1.87 (-4.9E-08) &	1.33 (-3.8E-08) & 0.76 (-2.0E-08) \\
5 &	2.28 (-1.1E-09) & 1.96 (-1.0E-08) & 1.46 (-2.9E-08) &	1.05 (-2.0E-08) & 0.60 (-9.9E-09) \\
10 &	1.69 (-8.4E-10) & 1.46 (-7.2E-09) & 1.09 (-2.0E-08) &	0.79 (-1.4E-08)	& 0.45 (-6.7E-09) \\
20 &	1.21 (-6.0E-10) & 1.04 (-5.1E-09) & 0.78 (-1.4E-08) &	0.56 (-9.6E-09) & 0.32 (-4.7E-09) \\ \bottomrule
\end{tabular}
\end{center}
\end{small}
\end{table}


\begin{table}
\caption{Prices and errors of the \citet{SP} method with Parameter Set 3
  in basis point units. Relative pricing errors, calculated as fractions of exact prices, are in parentheses.}
\label{tab10}
\begin{small}
\begin{center}
\begin{tabular}{cccccc}
\toprule
option &\multicolumn{5}{c}{swap maturity}\\
expiry & 1 & 2 & 5 & 10 & 30\\ \midrule

ATM & \multicolumn{5}{l}{$K = F$}\\
1 & 20.83 (1.8E-01) & 33.17 (2.6E-01) & 53.64 (3.7E-01) & 66.38 (4.3E-01)	& 71.32 (4.6E-01)\\
2 & 23.61 (1.4E-01) & 38.58 (2.0E-01) & 64.26 (2.7E-01) & 80.24 (3.2E-01)	& 86.41 (3.4E-01)\\
5 & 23.55 (1.0E-01) & 39.39 (1.4E-01) & 67.18 (1.9E-01) & 84.48 (2.2E-01)	& 91.14 (2.4E-01)\\
10 & 18.76 (7.4E-02) & 31.55 (1.0E-01) & 54.11 (1.4E-01) & 68.16 (1.6E-01) & 73.57 (1.7E-01)\\
20 & 10.90 (4.2E-02)	& 18.34 (5.8E-02) & 31.48 (7.8E-02) & 39.66  (9.1E-02) & 42.82 (9.7E-02)\\ \midrule

ITM & \multicolumn{5}{l}{$K = F -2\; \sigatm\sqrt{\Tex}$}\\
1 &	104.34 (3.3E-02) & 166.16 (5.1E-02) & 268.72 (8.6E-02) & 332.57 (1.3E-01) & 357.31 (2.0E-01) \\
2 &	118.56 (3.1E-02) & 193.75 (4.8E-02) & 322.73 (8.9E-02) & 403.00 (1.5E-01) & 433.98 (2.6E-01) \\
5 &	118.46 (2.7E-02) & 198.10 (4.4E-02) & 337.89 (8.9E-02) & 424.88 (1.7E-01) & 458.38 (3.1E-01) \\
10 &	94.40 (2.1E-02) & 158.75 (3.4E-02) &	 272.29 (7.2E-02) & 343.00 (1.4E-01) & 370.22 (2.6E-01) \\
20 &	54.85 (1.2E-02) & 92.30 (2.0E-02) & 158.45 (4.2E-02) & 199.64 (8.1E-02) & 215.50 (1.5E-01) \\ \midrule

OTM & \multicolumn{5}{l}{$K = F +2\; \sigatm\sqrt{\Tex}$}\\
1 &	0.46 (1.6E-02) & 0.73 (2.0E-02) & 1.17 (1.4E-02) & 1.45 (-1.5E-02) & 1.56 (-8.3E-02) \\
2 &	0.51 (7.4E-03) & 0.83 (5.8E-03) & 1.39 (-1.5E-02) & 1.73 (-6.8E-02) & 1.86 (-1.8E-01) \\
5 &	0.50 (8.7E-05) & 0.84 (-6.0E-03) & 1.44 (-3.9E-02) & 1.81 (-1.1E-01) &	 1.95 (-2.7E-01) \\
10 &	0.40 (-1.4E-03) & 0.67 (-7.3E-03) & 1.15 (-3.7E-02) & 1.45 (-1.0E-01) & 1.57 (-2.3E-01) \\
20 &	0.23 (-9.5E-04) & 0.39 (-4.4E-03) & 0.67 (-2.2E-02) &	0.85 (-5.9E-02) & 0.91 (-1.4E-01) \\ \bottomrule
\end{tabular}
\end{center}
\end{small}
\end{table}


\begin{table}
\caption{Implied normal volatilities and errors of the \citet{SP}
  method with Parameter Set 3 in daily basis point units. Relative
  volatility errors, calculated as fractions of exact volatilities,
  are in parentheses.}
\label{tab11}
\begin{small}
\begin{center}
\begin{tabular}{cccccc } \toprule

option &\multicolumn{5}{c}{swap maturity}\\
expiry &  1  &  2  & 5 & 10 & 30 \\ \midrule

ATM & \multicolumn{5}{l}{$K = F$}\\
1 & 3.62 (1.3E-02) & 2.96 (8.4E-03) & 2.08 (3.8E-03)	& 1.46 (2.3E-03) & 0.82 (1.2E-03)\\
2 & 3.07 (1.1E-02) & 2.57 (6.9E-03) & 1.86 (3.3E-03)	& 1.32 (2.0E-03) & 0.74 (1.1E-03) \\
5 & 2.28 (7.6E-03) & 1.96 (5.0E-03) & 1.45 (2.6E-03)	& 1.04 (1.7E-03) & 0.58 (9.3E-04) \\
10 & 1.69 (5.8E-03) & 1.46 (4.0E-03)	& 1.09 (2.2E-03)  & 0.78 (1.4E-03) & 0.44 (7.9E-04) \\
20 & 1.20 (4.4E-03) & 1.04 (3.0E-03)	& 0.77 (1.7E-03) & 0.55 (1.1E-03) & 0.31 (6.3E-04) \\ \midrule

ITM & \multicolumn{5}{l}{$K = F -2\; \sigatm\sqrt{\Tex}$}\\
1 &	3.62 (2.5E-02) & 2.96	 (1.9E-02) & 2.08 (1.4E-02 ) & 1.46 (1.4E-02) & 0.82 (1.4E-02) \\
2 &	3.07 (2.2E-02) & 2.57	 (1.7E-02) & 1.86 (1.5E-02) & 1.32 (1.5E-02) & 0.74 (1.5E-02) \\
5 &	2.28 (1.7E-02) & 1.96 (1.4E-02) & 1.45 (1.3E-02) & 1.04 (1.5E-02) & 0.58 (1.5E-02) \\
10 &	1.69 (1.3E-02) & 1.46 (1.1E-02) & 1.09 (1.0E-02) & 0.78 (1.2E-02) & 0.44 (1.2E-02) \\
20 &	1.20 (9.9E-03) & 1.04 (8.2E-03) & 0.77 (7.6E-03) & 0.55 (8.5E-03) & 0.31 (8.6E-03) \\ \midrule
OTM & \multicolumn{5}{l}{$K = F +2\; \sigatm\sqrt{\Tex}$}\\
1 &	3.62 (1.7E-03) & 2.96 (-1.9E-03) & 2.08 (-6.6E-03) & 1.46 (-9.7E-03) &	 0.82 (-1.1E-02) \\
2 &	3.07 (-6.4E-04) & 2.57 (-3.6E-03) & 1.86 (-7.9E-03) & 1.32 (-1.1E-02) & 0.74 (-1.3E-02) \\
5 &	2.28 (-2.2E-03) & 1.96 (-4.2E-03) & 1.45 (-7.7E-03) &	1.04 (-1.1E-02) & 0.58 (-1.3E-02) \\
10 &	1.69 (-1.8E-03) & 1.46 (-3.2E-03) & 1.09 (-5.9E-03) &	0.78 (-8.7E-03) & 0.44 (-1.0E-02) \\
20 &	1.20 (-1.1E-03) & 1.04 (-2.1E-03) & 0.77 (-4.1E-03) &	0.55 (-6.1E-03) & 0.31 (-7.2E-03) \\ \bottomrule
\end{tabular}
\end{center}
\end{small}
\end{table}

The swaption pricing errors are shown in Tables~\ref{tab4} to
\ref{tab11}. For each example, we first present the price and its
error in basis points for a $5\times5$ swaption matrix. Then, we
provide the implied normal volatility and its error. The normal
volatility is the volatility under the Bachelier process, i.e., normal
diffusion. For our study, we assume that the normal volatility is more
relevant than the Black-Scholes (or log-normal) volatility. First, the
normal volatility is widely used among practitioners in the fixed
income area~\citep{Choi}. Second, the short rate or IFR in the GTSM
follows the Bachelier process, and the same holds nearly true for the
swap rate (in fact, this is the key assumption of \citet{SP}, and we
will discuss its accuracy shortly).  Therefore, the normal volatility
is nearly constant across options with different strikes, which makes
it a better measure of error than the price. The price of options can
change drastically as moneyness changes; thus, pricing errors, both
relative and absolute, can be misleading, whereas the normal
volatility is a consistent measure of error regardless of the
moneyness.

We further convert the normal volatility to daily basis point (DBP)
units by multiplying it by $10^4/\sqrt{252}$, assuming that there are
252 trading days in a year. The DBP volatility offers an intuitive
measure of the average daily change in the underlying swap rate.

In each table, we use three different strikes: at-the-money (ATM),
out-of-the-money (OTM) and in-the-money (ITM). To maintain the
consistent moneyness of the OTM and ITM options across the surface, we
use
\begin{equation}
  K = F \pm n\;\sigatm\;\sqrt{\Tex} \quad\text{for}\quad n= 0.5,\;
1,\;\text{or}\; 2
\end{equation}
where $F$ is the forward swap rate, and $\sigatm$ is the normal
volatility for ATM.

The accuracy of the hyperplane approximation is uniformly good across
the volatility surface for all three examples. The maximum volatility
error across all examples is of the order of $10^{-6}$ DBP. This level
of error does not require further correction for practical purposes.

In particular, our method gives results superior to those from the
method of \citet{SP} because it accurately captures the skew in the
normal volatility.  For comparison, we reproduce the results of
\citet{SP} for Parameter Set 3; compare Tables~\ref{tab10} and
\ref{tab11} to Tables~\ref{tab8} and \ref{tab9}. The error in
\citet{SP}'s method is primarily caused by the condition that the
normal implied volatility is constant across strikes, whereas the GTSM
has a slightly upward sloping volatility skew, as indicated by our
hyperplane approximation and exact methods. This tendency arises
because LVMs are assumed to be constant at their time-zero values in
deriving the SDE for the swap rate in \citet{SP}.  It should be
mentioned that \citet{piterbarg_vol2_swapdyn} further refine the swap
rate SDE in the broader context of the linear local volatility
Gaussian model. In their improved SDE, the swap rate follows a
displaced log-normal diffusion, thus exhibiting the volatility skew.
We do not implement their method here and leave the performance
comparison for future study.

\begin{figure}[!h]
\centering
\subfigure[Exact exercise boundary]{\includegraphics[width=0.8\textwidth]{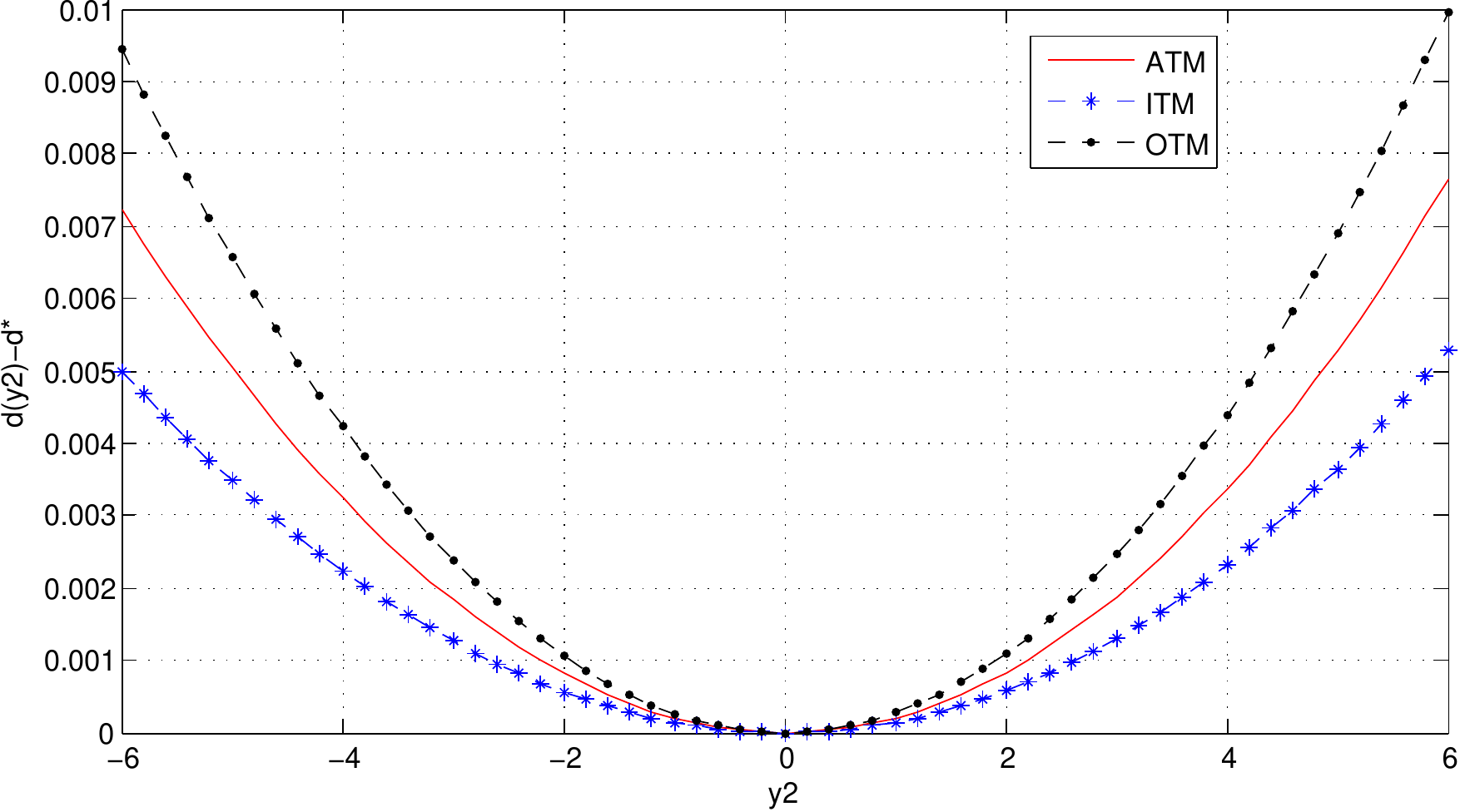}}\\
\subfigure[Error density]{\includegraphics[width=0.8\textwidth]{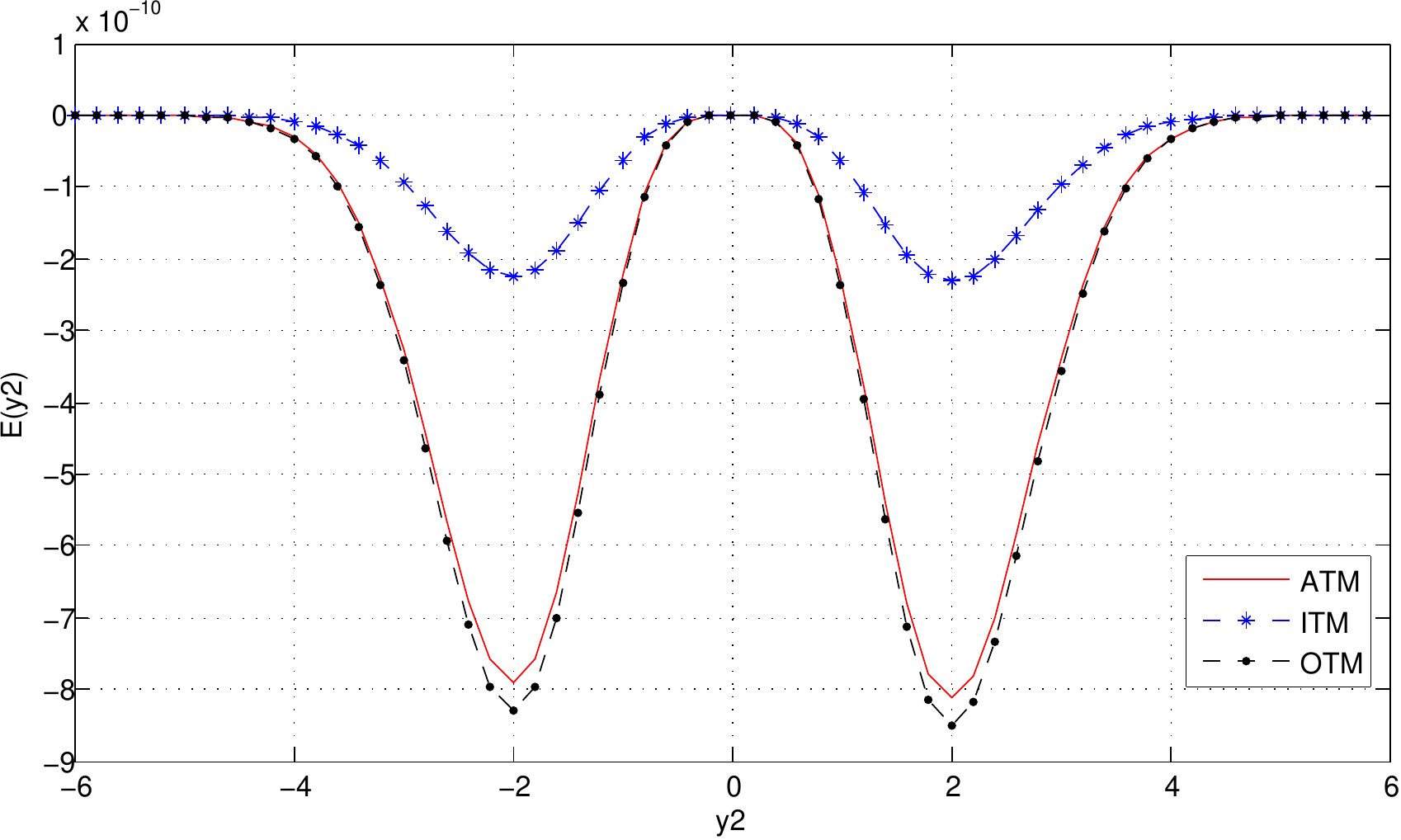}}

\caption{(a) The exact exercise boundary of the 2 $\times$ 10
  swaptions in Parameter Set 1. The strikes for ATM, ITM and OTM are
  5.06\%, 3.51\% and 6.62\%, respectively. Our method approximates
  this boundary as the horizontal axis $d(y_2)=d^*$. Both the $x$ and
  $y$ axes are normalized by the standard deviation of the state
  variables.  Although the distance between the exact boundary and the
  hyperplane grows quadratically from the tangent point, the
  probability of the normal distribution decays significantly
  faster. (b) The error density defined in Eq.~(\ref{eq:error}) for
  the same swaptions and parameter sets. It is the
  probability-weighted swaption payoff integrated over the area
  between the exact boundary $\partial \Omega$ and the approximated
  hyperplane $\partial \tilde{\Omega}$ (shaded area in
  Fig.~\ref{fig:hyperplane}) in the direction of $y_1$. The error
  peaks at approximately two standard deviations and quickly decays
  because of the normally distributed probability density.}
\label{fig:boundary}
\end{figure}

\begin{figure}[!h]
\centering
\subfigure{\includegraphics[width=0.7\linewidth]{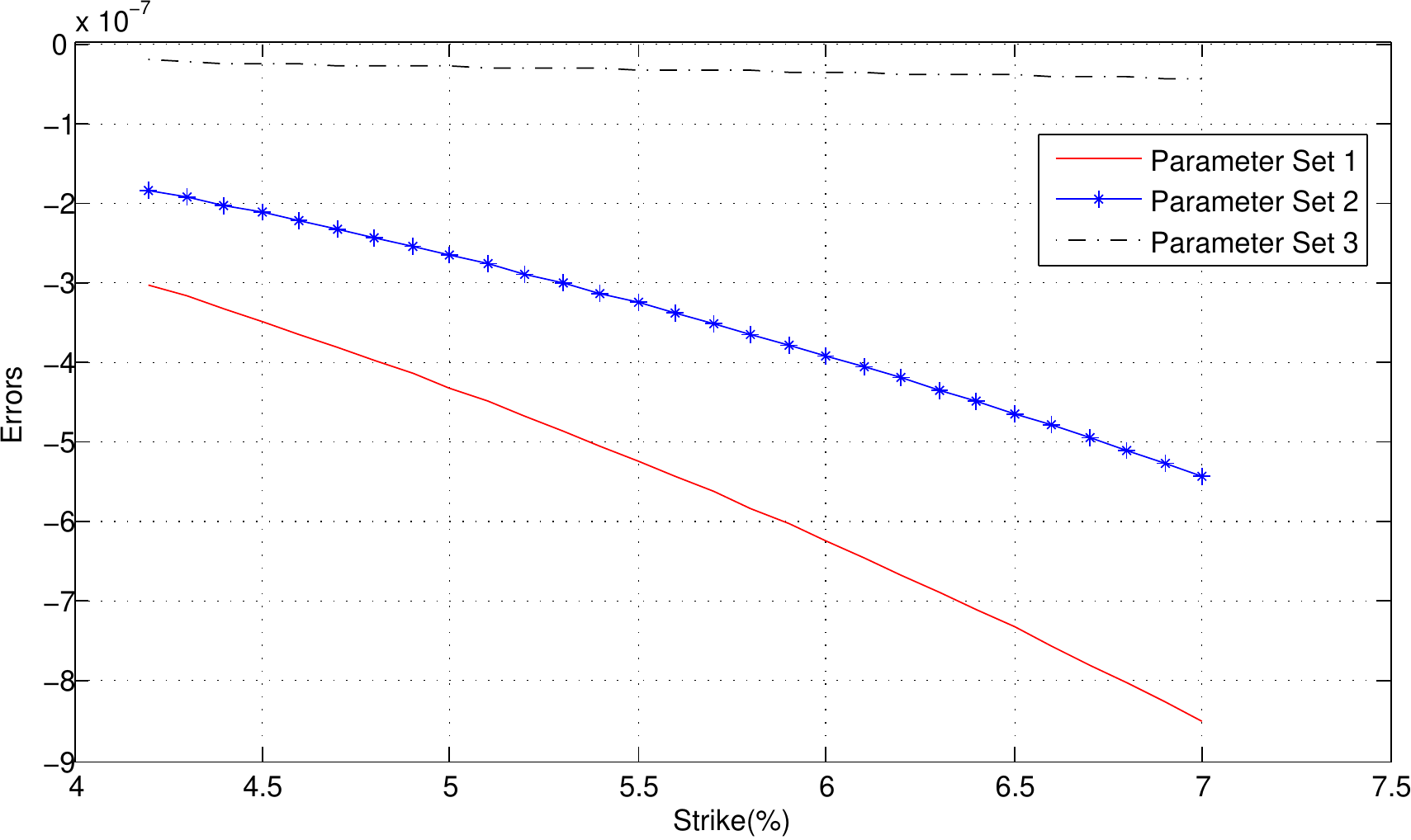}}
\caption{The implied volatility errors of the hyperplane approximation
  for varying strikes in daily basis point units. We use a 2 $\times$
  10 payer swaption against the three parameter sets. Although the
  error increases as the swaption becomes more out-of-the money, the
  approximation results remain good for very high strikes.}
\label{fig:error}
\end{figure}

We further analyze the error using a particular example: a $2\times10$
swaption on Parameter Set 1. First, we present the exact exercise
boundary for this case in Fig.~\ref{fig:boundary}(a). The boundary
lines for different strikes are slightly convex upward but are close
to flat lines.  In our method, by approximating the boundary with a
flat line, we incorrectly exercise the swaption when the state falls
into the area between the boundaries where the underlying swap has a
negative value. Thus, we have a negative pricing error of $-2.8\times
10^{-5}$ for ATM. In Fig.~\ref{fig:boundary}(b), we provide the error
density as defined in Eq.~(\ref{eq:error}). Finally, we plot the DBP
volatility error as a function of the strike in
Fig.~\ref{fig:error}. For all three parameter sets, the error tends to
increase for a higher strike.  This increase is most likely because
each term in Eq.~(\ref{eq:recon2}) becomes more convex as the state
becomes larger; this increases the deviation between the exercise
boundary and the flat line.

\bibliographystyle{plainnat}
\bibliography{fastcalib}

\begin{thebibliography}{14}
\providecommand{\natexlab}[1]{#1}
\providecommand{\url}[1]{\texttt{#1}}
\expandafter\ifx\csname urlstyle\endcsname\relax
  \providecommand{\doi}[1]{doi: #1}\else
  \providecommand{\doi}{doi: \begingroup \urlstyle{rm}\Url}\fi

\bibitem[Andersen and Piterbarg(2010{\natexlab{a}})]{piterbarg_vol2}
Leif~{B.G.} Andersen and Vladimir~V. Piterbarg.
\newblock \emph{Interest Rate Modeling. Volume 2: Term Structure Models},
  chapter 10.2.5.
\newblock Atlantic Financial Press, August 2010{\natexlab{a}}.
\newblock ISBN 0984422110.

\bibitem[Andersen and Piterbarg(2010{\natexlab{b}})]{piterbarg_vol2_corr}
Leif~{B.G.} Andersen and Vladimir~V. Piterbarg.
\newblock \emph{Interest Rate Modeling. Volume 2: Term Structure Models},
  chapter 12.1.3.
\newblock Atlantic Financial Press, August 2010{\natexlab{b}}.
\newblock ISBN 0984422110.

\bibitem[Andersen and Piterbarg(2010{\natexlab{c}})]{piterbarg_vol2_swapdyn}
Leif~{B.G.} Andersen and Vladimir~V. Piterbarg.
\newblock \emph{Interest Rate Modeling. Volume 2: Term Structure Models},
  chapter 13.1.5.
\newblock Atlantic Financial Press, August 2010{\natexlab{c}}.
\newblock ISBN 0984422110.

\bibitem[Bakshi and Madan(2000)]{MB}
Gurdip Bakshi and Dilip Madan.
\newblock Spanning and derivative-security valuation.
\newblock \emph{Journal of Financial Economics}, 55\penalty0 (2):\penalty0
  205--238, February 2000.

\bibitem[Choi et~al.(2009)Choi, Kim, and Kwak]{Choi}
Jaehyuk Choi, Kwangmoon Kim, and Minsuk Kwak.
\newblock Numerical approximation of the implied volatility under arithmetic
  brownian motion.
\newblock \emph{Applied Mathematical Finance}, 16\penalty0 (3):\penalty0
  261--268, 2009.
\newblock \doi{10.1080/13504860802583436}.

\bibitem[{Collin-Dufresne} and Goldstein(2002)]{CDP}
Pierre {Collin-Dufresne} and Robert~S. Goldstein.
\newblock Pricing swaptions within an affine framework.
\newblock \emph{Journal of Derivatives}, 10\penalty0 (1):\penalty0 9--26, 2002.

\bibitem[Duffie et~al.(2000)Duffie, Pan, and Singleton]{DPS}
Darrell Duffie, Jun Pan, and Kenneth~J Singleton.
\newblock Transform analysis and asset pricing for affine jump-diffusions.
\newblock \emph{Econometrica}, 68\penalty0 (6):\penalty0 1343--1376, November
  2000.

\bibitem[Heath et~al.(1992)Heath, Jarrow, and Morton]{HJM}
David Heath, Robert~A. Jarrow, and Andrew Morton.
\newblock Bond pricing and the term structure of interest rates: A new
  methodology for contingent claims valuation.
\newblock \emph{Econometrica}, 60\penalty0 (1):\penalty0 77--105, 1992.

\bibitem[Ho and Lee(1986)]{HoLee}
Thomas S~Y Ho and Sang-bin Lee.
\newblock Term structure movements and pricing interest rate contingent claims.
\newblock \emph{Journal of Finance}, 41\penalty0 (5):\penalty0 1011--29,
  December 1986.

\bibitem[Hull and White(1990)]{Hull}
John Hull and Alan White.
\newblock Pricing interest-rate-derivative securities.
\newblock \emph{Review of Financial Studies}, 3\penalty0 (4):\penalty0 573--92,
  1990.

\bibitem[Munk(1999)]{Munk}
Claus Munk.
\newblock Stochastic duration and fast coupon bond option pricing in
  multi-factor models.
\newblock \emph{Review of Derivatives Research}, 3\penalty0 (2):\penalty0
  157--181, 1999.
\newblock ISSN 1380-6645.

\bibitem[Schrager and Pelsser(2006)]{SP}
David~F Schrager and Antoon A.~J Pelsser.
\newblock Pricing swaptions and coupon bond options in affine term structure
  models.
\newblock \emph{Mathematical Finance}, 16\penalty0 (4):\penalty0 673--694,
  October 2006.
\newblock ISSN 1467-9965.

\bibitem[Singleton and Umantsev(2002)]{SU}
Kenneth~J Singleton and Len Umantsev.
\newblock Pricing coupon-bond options and swaptions in affine term structure
  models.
\newblock \emph{Mathematical Finance}, 12\penalty0 (4):\penalty0 427--446,
  October 2002.
\newblock ISSN 1467-9965.

\bibitem[Vasicek(1977)]{Vasicek}
Oldrich Vasicek.
\newblock An equilibrium characterization of the term structure.
\newblock \emph{Journal of Financial Economics}, 5\penalty0 (2):\penalty0
  177--188, November 1977.

\end{thebibliography}

\end{document}